\documentclass[utf8,twocolumn]{jetpe} 

\usepackage[russian]{babel}
\usepackage{cite}
\usepackage{color}
\usepackage{url}
\usepackage{aas_macros}

\title{On the increase in the concentration of primordial black holes in the halos of dwarf galaxies}
\rtitle{PBH impact on DM halos}

\setaffiliation1{Astro Space Center of P.N. Lebedev Physical Institute of the RAS, Profsoyuznaja 84/32, 117997 Moscow, Russia}

\setauthor{S.~V.}{Pilipenko}{1}
\email{spilipenko@asc.rssi.ru}

\setauthor{M.~V.}{Tkachev}{1}
\email{mtkachev@asc.rssi.ru}

\setauthor{N.~R.}{Arakelyan}{1}

\rauthor{Pilipenko, Tkachev, Arakelyan}

\abstract{
Through numerical experiments, we have predicted that if dark matter (DM) contains even a small fraction, $f_0\sim10^{-4}$, of primordial black holes (PBHs), during the formation of the gravitationally bound halo of a dwarf galaxy, these PBHs will concentrate in a region with a radius of about 10~pc, so that their local fraction will exceed 1\%. Unlike previous studies of PBH migration to the centers of galaxies, the numerical experiments conducted here take into account the early formation of a massive ``dress'' of DM around the PBHs and the non-stationarity of the halo during its formation. Applying our results to models of heating stellar clusters in the Eridanus II and Segue I galaxies due to dynamical friction between stars and PBHs allows us to impose constraints on the abundance of PBHs that are two orders of magnitude stricter than previously thought.
}

\begin{document}

\maketitle 


\textbf{1. Introduction.}
The nature of dark matter (DM) is still unclear, and one of its components could be primordial black holes (PBHs) -- objects formed in the early Universe, long before the formation of the first stars \cite{ZN67}. Current constraints converge on the idea that, apparently, dark matter cannot consist solely of PBHs, however, observations of gravitational waves from mergers of relatively high-mass black holes (greater than 60 solar masses) are viewed by some researchers as evidence supporting the presence of PBHs, see, for example, \cite{Carr21}. Some inflationary theories predict the formation of PBHs, so their observational detection or exclusion is an important test for models of the early Universe, see, for example, \cite{Ivanov94}.

One manifestation of PBHs with masses greater than the solar mass is their impact on the dynamics of gravitationally bound systems, particularly dwarf galaxies. If PBHs are more massive than stars, they will heat the stellar system due to dynamical friction. Based on observational data on the structure of the dwarf galaxies Eridanus II and Segue I, this effect has been used in the literature to impose constraints on the abundance of PBHs \cite{Brandt16,Zhu18,2020A&A...635A.107Z,Graham24a,Koulen24,Belotsky24} (in the latter work, the mass of PBHs is less than the stellar mass, but they are grouped into more massive clusters, which are also subject to dynamical friction). In particular, the center of the Eridanus II galaxy hosts an old star cluster, which would be destroyed if the fraction of PBHs exceeds $10^{-3}$ and the mass of a single PBH is $10^3$~M$_\odot$ \cite{Koulen24}. Several studies have also explored the interaction of PBHs with dark matter composed of particles (PDM): due to the same dynamical friction, PBHs condense toward the center of the halo \cite{Koulen24}, and in the distribution of the PDM itself the ``cusp'' -- a region near the center of the halo where the density in spherically symmetric shells behaves approximately as $r^{-1}$ -- is destroyed \cite{Boldrini20}. The work \cite{Graham24a} particularly emphasizes the importance of the condensation of PBHs toward the halo center, as this significantly strengthens the obtained constraints on their fraction $f$. Several cosmological numerical simulations of large-scale structure formation, including PBHs and DM, have been performed \cite{Inman_19,Tkachev20,Tkachev22}; these have allowed the establishment of general patterns of structure growth and provided constraints on the merger rate of binary PBHs producing gravitational wave signals, while the resolution of these models is insufficient for studying the internal structure of dwarf galaxies. The works \cite{Stasenko23,Stasenko24} consider the formation of PBH halos and clusters due to Poissonian perturbations induced by PBHs and study the migration of PBHs within halos under the influence of dynamical friction.

The existing literature on the interaction of PBHs with dark matter (DM) in the halos of dwarf galaxies does not take into account two important factors: first, isolated PBHs in the early Universe create perturbations in the gravitational potential, which lead to the formation of mini-halos (or ``dresses'') of DM around each PBH \cite{MOR_2007,Eroshenko_16}. By the time the halos of galaxies begin to form, these mini-halos should have masses 100 times greater than the mass of the PBHs themselves and compact sizes, so they should influence the further evolution of PBHs during structure formation and especially during PBH mergers (see, for example, \cite{Inman_19,Pilipenko22}). Second, in studies of PBH dynamics within DM halos \cite{Zhu18,Boldrini20,Graham24a,Koulen24,Stasenko23,Stasenko24}, the simplest distribution function for DM particles was used: the halo was assumed to be spherically symmetric and stationary, and the velocity distribution was assumed to be isotropic. This significantly differs from the structure of DM halos obtained in cosmological numerical models, where they form from small density perturbations due to gravitational instability. Such halos are significantly non-spherical, and in the early stages of their existence, they are non-stationary, with the velocity distribution being highly anisotropic and demonstrating the presence of numerous separate streams of DM at each point in three-dimensional space \cite{Navarro10,Kuhlen10,Vogelsberger11}. Therefore, we consider it necessary to refine the current results by using a more realistic model of the halo.

In this work, we model the process of halo formation in a dwarf galaxy, consisting of a mixture of PDM and PBHs, starting from the moment before the Universe transitions from the radiation-dominated stage to the matter-dominated stage (Section 2). This allows us to track both the formation of minihalos of DM around each PBH and the formation of the main halo due to the development of large-scale density waves. Meanwhile, PBHs, surrounded by their minihalos, interact with each other and with the streams of PDM, which leads to a significant increase in the density of PBHs at the centers of the halos (Section 3). This allows us to tighten the existing constraints obtained earlier, since the local fraction of PBHs in the DM content at the centers of dwarf galaxies is now significantly higher than the average across the Universe (Section 4).

\textbf{2. Simulations of halo formation.}
As a time measure in cosmological numerical calculations, it is customary to use the scale factor $a$, which is equal to 1 at the present time. It is related to time through cosmological parameters. We assume the Universe to be flat and use the standard parameters provided by the Planck collaboration \cite{Planck18_VI}.

Our numerical models are constructed in such a way as to reproduce two processes: the formation of a minihalo around a PBH, which begins in the early Universe when $a<10^{-3}$ ($z>1000$), as well as the formation of a halo of a dwarf galaxy, which occurs when $a\sim0.1$ ($z\sim10$). The halo then evolves for several billion years until it reaches a state close to the present ($a\sim1$). To account for the ``infinite'' Universe, periodic boundary conditions are applied, and the calculations are performed in comoving cosmological coordinates using the GADGET-2 code \cite{gadget}, which we modified to account for a uniformly distributed radiation component, which is important when $a<10^{-3}$. Also we use a GPU version of GADGET-2\footnote{\url{https://github.com/pseudotensor/cuda-gadget}}.

As shown in \cite{1985ApJS...58...39B,MOR_2007,Eroshenko_16,Pilipenko22}, during the radiation-dominated era of the Universe, when $a<3\times10^{-4}$ ($z>3300$), a halo of PDM begins to form around an isolated PBH, but this growth is negligible (its mass is on the order of the mass of the PBH itself). In the matter-dominated era, the growth continues and follows the law 
\begin{equation} 
\label{eq:halo_growth} 
M_H \approx m_{PBH} \frac{a}{a_{eq}}, 
\end{equation} 
where $M_H$ is the mass of the minihalo within the {turnaround} radius at which the cosmological expansion of matter turns into contraction, $m_{PBH}$ is the PBH mass, and $a_{eq}$ is the scale factor at the time of matter-radiation equality.

\begin{figure} \centering \includegraphics[width=\linewidth]{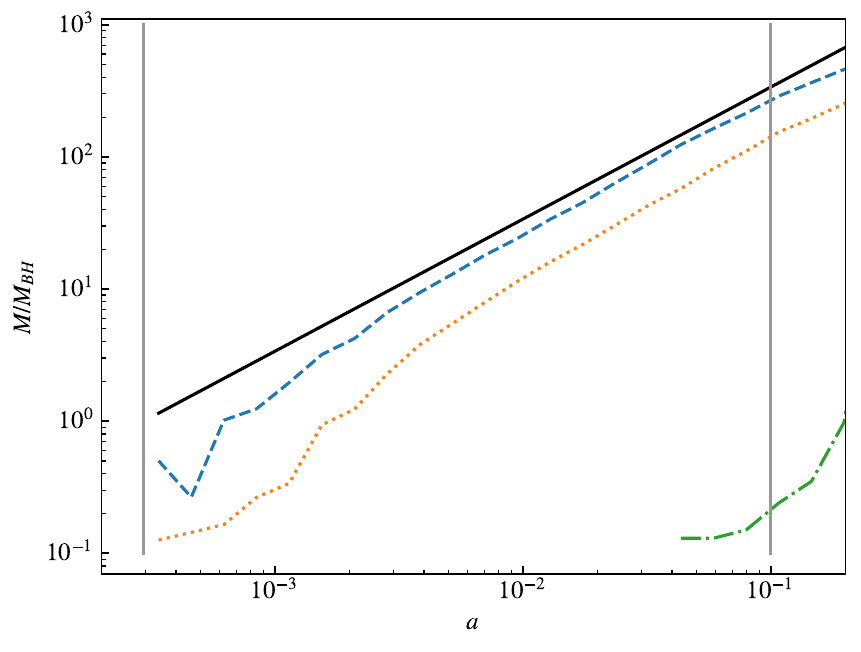} \caption{Dependence of the average minihalo mass in units of the PBH mass on the scale factor for models where the initial conditions are set at $a_\mathrm{init} = 3\times10^{-5}$ (dashed line) and $a_\mathrm{init} = 3\times10^{-4}$ (dotted line), $a_\mathrm{init} = 4.3\times10^{-2}$ (dash-dotted line). The solid line shows equation (\ref{eq:halo_growth}). Vertical lines indicate the moments $a_{eq}$ (left) and the moment of halo formation (right).} \label{fig:minihalo_growth} \end{figure}

It should be noted that due to mass conservation, a certain void must have formed around the PBH as a result of the material that went into the creation of the PBH itself. On the other hand, the perturbations with enhanced density, from which the PBHs originated, likely extended beyond the region that collapsed into the PBH. This means that from the very beginning, there must have been some perturbations, both with enhanced and reduced density, around the PBH, which should affect the growth of the minihalo around the PBH, at least at early stages. We do not account for these features since they depend on the details of the PBH formation model, but by varying the simulation start time, we obtain minihalos with different masses at the time of the formation of the main halo. Also, in many PBH models, they form in clusters (see, for example, \cite{Belotsky19}), but for simplicity, we do not consider this clustering in this work.

{As is customary in cosmological simulations, the environment of non-interacting PDM is modeled using a finite number of test particles with mass $m_{DM}$ much larger than the mass of the PDM particles themselves, in our case $m_{DM}=5$~M$_\odot$.} To set the initial conditions, we place {test} DM particles on a regular Cartesian grid. Then, at some moment $a_\mathrm{init}$, we replace a certain number of randomly selected particles with more massive particles representing PBHs. The mass of the PDM particles is then reduced to maintain the average cosmological DM density. The number of particles (PDM + PBH) in all calculations was $2^{24}\approx16.7\times10^6$, corresponding to 256 particles along one side of the simulation box in the initial distribution on the Cartesian grid. The number of PBHs was chosen so that the mass fraction of PBHs relative to the total DM mass was $f_0=10^{-3}$, $f_0=3\times10^{-4}$, or $f_0=10^{-4}$.

To account for possible differences in the mass of the minihalo from the predictions of formula (\ref{eq:halo_growth}), we vary the initial moment $a_\mathrm{init}$. Figure~\ref{fig:minihalo_growth} shows three variants of the minihalo growth history. One of them, with $a_\mathrm{init}=0.043$ ($z=22$), {was added by us to illustrate} the system without a minihalo at the moment of the formation of the galactic halo. The second option $a_\mathrm{init}=3\times10^{-5}$ ($z=33000$) approximately reproduces formula (\ref{eq:halo_growth}), while the third, with $a_\mathrm{init}=3\times10^{-4}$ ($z=3300$), gives a minihalo mass approximately 3 times smaller than the prediction of formula (\ref{eq:halo_growth}).

Note that the medium with randomly located PBHs is inhomogeneous, including on scales much larger than the average distance between PBHs, which leads over time to the development of large-scale structures from PDM, such as filaments \cite{Afshordi03,Inman_19}. In such structures, the accretion rate of matter onto the minihalo differs from the accretion rate onto a minihalo in a homogeneous background, and therefore, the masses of individual minihalos also differ from the average value shown in Figure~\ref{fig:minihalo_growth}. 

We considered three variants of the PBH to DM particle mass ratio: $m_{PBH}/m_{DM} = 20,$ 100, 1000, which allowed us to investigate the effect of the PBH mass on the result. The growth of the minihalo mass, expressed in PBH masses, occurs in almost the same way for these three variants.

\begin{figure} \centering \includegraphics[width=0.5\textwidth]{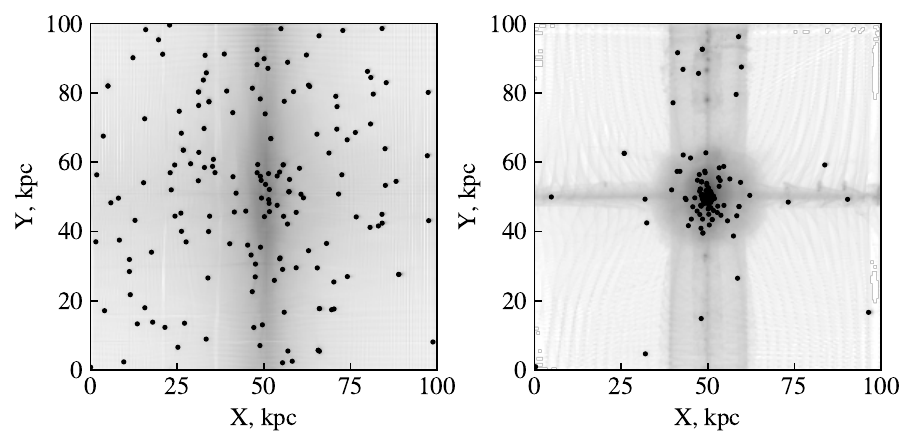} \caption{Evolution of the matter distribution in the simulation box in comoving coordinates, projected onto the XY plane. Left: at $a=0.043$, right: at $a=0.5$. The PDM density is shown in gray, and the PBHs are shown as black dots.} \label{fig:xy} \end{figure}

\begin{figure} \centering \includegraphics[width=0.5\textwidth]{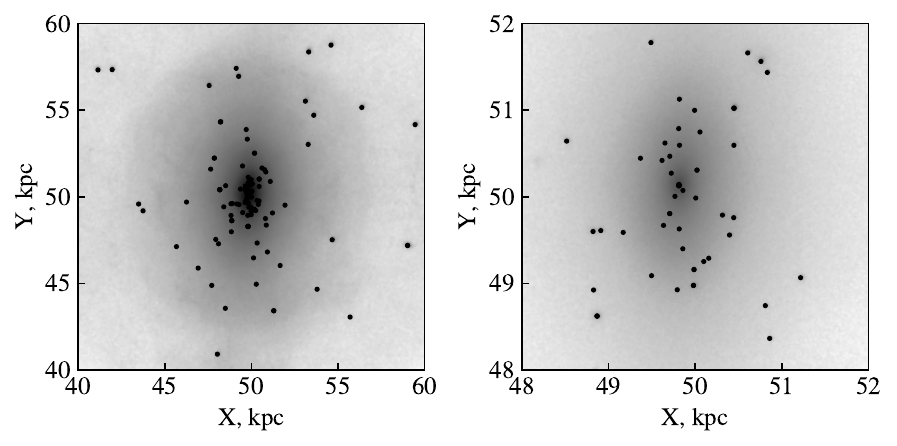} \caption{Enlarged region from the right panel of Figure~\ref{fig:xy}, showing the dark matter halo. The PDM density is shown in gray, and the PBHs are shown as black dots.} \label{fig:xy_center} \end{figure}

To form a dwarf galaxy halo, we set three large-scale density waves along the axes of the simulation box. The amplitudes of the waves are in a ratio of 10:4:3 along the X, Y, and Z axes of the simulation box, which allows the halo to be sufficiently anisotropic. {This amplitude ratio is chosen as an example and is quite typical; the distribution function of amplitudes is given, for example, in \cite{Doroshkevich70}.} Specifically, we set the particle displacements and velocities according to the expressions: 
\begin{eqnarray}\label{eq:init} 
& \Delta x_{i_k} = A_k \mathrm{sin}\left( \frac{2\pi i_k}{n_1} \right),\nonumber \\ 
& v_{i_k} = A_k V(a) \mathrm{sin}\left( \frac{2\pi i_k}{n_1} \right), 
\end{eqnarray} 
where $A_k$ are the perturbation amplitudes along the three axes, $V(a)$ is the time-dependent velocity amplitude, $i_k$ is the particle number along the $k$-th axis in the Cartesian grid, and $n_1=256$ is the number of particles along one axis. These waves are related to the 
primordial density perturbations, not to Poissonian perturbations imposed by PBHs, which, for the considered fractions $f_0<10^{-3}$, are too weak to form a significant number of structures by $z\sim10$.

Our model resembles the halo formation scenario in cosmological numerical simulations, where perturbations of different scales with random amplitudes and phases are present. As a rule, along one of the directions, the perturbation amplitude is the largest, so a flat ``pancake'' \cite{Zeldovich70} forms, within which a halo forms \cite{Valinia97}. Similar initial conditions were used in the literature \cite{Shapiro04,Pilipenko12}, and it was shown that as a result, a halo with an NFW density profile \cite{NFW} forms, as in large-volume cosmological simulations. The wave amplitude along the X-axis is chosen so that the ``pancake'' forms at $a=0.08$ ($z=11.5$). At the same moment, thanks to the waves along the Y and Z axes, a density enhancement begins to grow at the center of the simulation box, forming a dwarf galaxy halo. By the time $a=0.1$ ($z=9$), 20 million years after the start of formation, the virial mass of the halo reaches 20\% of the mass of the box, and the halo growth significantly slows down. By $a=0.5$ ($z=1$, the end of the simulation, 5.4 billion years after the start of halo formation), the halo mass reaches 50\% of the mass of the box. This growth history resembles the formation of present-day dwarf galaxies \cite{BarkanaLoeb,Demianski23}. It is also worth noting that the dynamical time at the center of the halo is about 5 million years.

For convenience of comparison with observations, we set the comoving size of the simulation box as $L_{box}=0.1$ Mpc. Our results can be easily rescaled with the box size: the results do not change if this size is multiplied by $s$, all masses by $s^{3}$, and velocities by $s$. In particular, for $s=1$, the virial mass of the halo at the end of the simulation reaches $5\times10^7$~M$_\odot$, and the mass of one PBH is 100, 500, or 5000 $M_\odot$ for different calculation variants.

An important characteristic of numerical calculations with dark matter is the gravitational potential smoothing length, which in our calculations was 1~pc. This allows us to resolve structures of sizes from several parsecs. We also checked the effect of smoothing by making two additional calculations for one of the modeled systems: one with twice the smoothing length and the other with half the smoothing length. The results did not change within numerical noise, indicating that the smoothing length was large enough.

We performed calculations for three values of the PBH mass fraction: $f_0=10^{-3}$, $f_0=3\times10^{-4}$, and $f_0=10^{-4}$. For $f_0=10^{-3}$, the PBH mass $m_{PBH}/m_{DM}$ was set to 20, 100, and 1000 times the test particle DM mass, as described above. However, for $f_0=3\times10^{-4}$ and smaller, only the 20 and 100 variants were considered, since with $m_{PBH}=1000 m_{DM}$ and such a low fraction, on average, only 1.7 PBHs fit into the simulation box. In our calculations, the number of PBHs in the simulation box varied from 17 to 850. An example of how our simulations look is shown in Figures~\ref{fig:xy} and \ref{fig:xy_center} for the case $f_0=10^{-3}$, $m_{PBH}/m_{DM}=500$. The minihalos around PBHs in these figures are almost invisible due to their small size compared to the black dots representing the PBHs. Figures~\ref{fig:xy} and \ref{fig:xy_center} are shown in comoving coordinates, in which the calculations were performed. All subsequent figures in this paper concern the internal structure of the halo and are presented in physical coordinates obtained from comoving coordinates by multiplying by the scale factor $a$.

It should be noted that in our calculations, there is no need to account for the effects of General Relativity, in particular, PBH mergers, since according to existing observational constraints on gravitational waves, only a few percent of dwarf galaxies would have experienced a single merger during the lifetime of the Universe. This estimate is made as follows, based on the Press-Schechter mass function: the density of halos with mass $5\times10^7$ (in a logarithmic mass interval) is about 30 per Mpc$^3$. These halos account for no more than 10\% of the total halo mass, i.e., they contribute to about 10\% of all PBH mergers. According to observations, the merger rate is 10-100 events per Gpc$^3$/year. This corresponds to $0.03-0.3$ events over 10 billion years per halo of the indicated mass. {We also do not account for the influence of the baryonic component on PBHs. The evolution of the baryonic component in galaxies is extremely complex, but in the future, it may become possible to model {both the PBHs and} the baryonic component. The objects considered, dwarf spheroidal galaxies, currently have an extremely low fraction of gas and stars compared to dark matter, meaning that at least in the later stages of the system's evolution, considering baryons is not necessary. This distinguishes them from large galaxies, where the interaction of PBHs with stars must always be considered.}

{We assume cold PDM, namely, we assume that the galaxy halo with mass $M$ formed in an medium without pressure from density waves with a wavelength of about $(M/\rho_m)^{1/3}$, where $\rho_m$ is the average matter density. This is also true in models with warm DM if the DM particle mass $m_{WDM}>10/s$~keV. Otherwise, halos of dwarf galaxies form by fragmentation of much larger structures made of dark matter, see, for example, \cite{Angulo13}.}

\begin{figure}
    \centering
    \includegraphics[width=\linewidth]{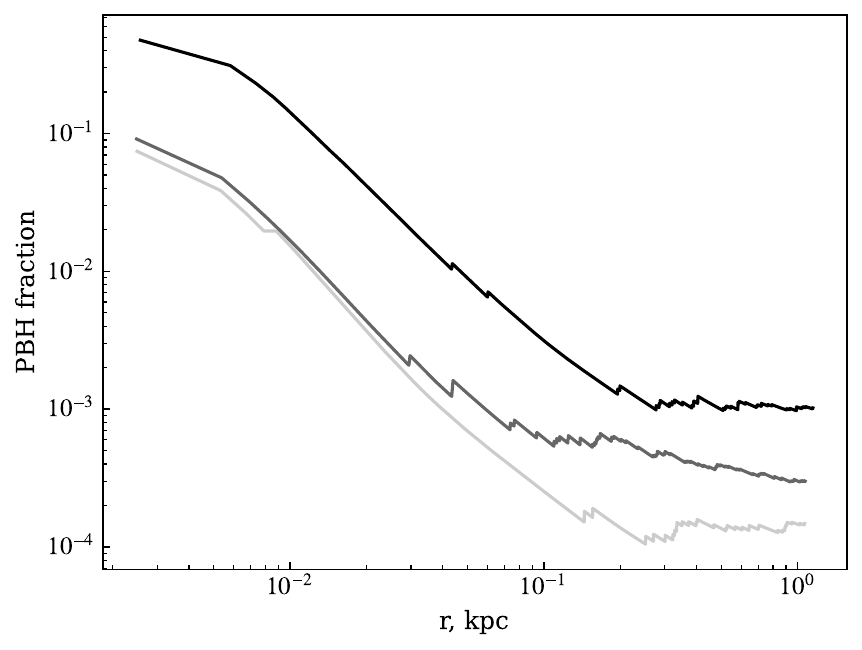}
    \caption{PBH fraction in a sphere of radius $ r $ for three values of the cosmological PBH fraction $ f_0 = 10^{-4} $, $ 3 \times 10^{-4} $, and $ 10^{-3} $ (from light to dark as $ f_0 $ increases) at $ a = 0.5 $.}
    \label{fig:fraction}
\end{figure}

\textbf{3. Condensation of PBHs in the halo center.}
Our main result is the detection of a significant increase in the PBH fraction after the formation of the galaxy's halo at its center. We define the local fraction $ f(r) $ as the ratio of the total PBH mass within a sphere of radius $ r $ to the total DM mass (PBH + PDM) in the same sphere. From Figure~\ref{fig:fraction}, we can see that at $ f_0 = 10^{-4} $, at a distance of 10~pc from the galaxy center, the fraction $ f $ reaches 1\%, i.e., it increases by 100 times, and at $ f_0 = 10^{-3} $, the fraction at the center reaches 10\%. This result is given for $ m_{PBH} = 100 \, M_\odot $ ($ m_{PBH}/m_{DM} = 20 $).

The coincidence of the fraction in the center for $ f_0 = 10^{-4} $ and $ 3 \times 10^{-4} $ in Figure~\ref{fig:fraction} is due to the small number of PBHs in the center of the halo for $ r < 10 $~kpc: in the model with $ f_0 = 10^{-4} $, there are only 4 PBHs, and in the model with $ f_0 = 3 \times 10^{-4} $, there are 5 PBHs at this distance, i.e., the dispersion in the number of PBHs is expected to be of the same order as the measured number of PBHs. In the model with $ f_0 = 10^{-3} $, there are already 22 PBHs in the center. We approximate the data from Figure~\ref{fig:fraction} with the relation $ f(r < 10 \text{pc}) = 43 f_0 $, which gives the minimum value of the reduced chi-square statistic: $ \chi^2 = 0.28 $\footnote{{The measurement error is estimated as $ f_m(r)/\sqrt{N_{PBH}-1} $, where $ f_m $ is the measured fraction.}}. Therefore, we consider that the PBH fraction in the center of the halo in our numerical models is proportional to the initial fraction $ f_0 $.

\begin{figure}
    \centering
    \includegraphics[width=\linewidth]{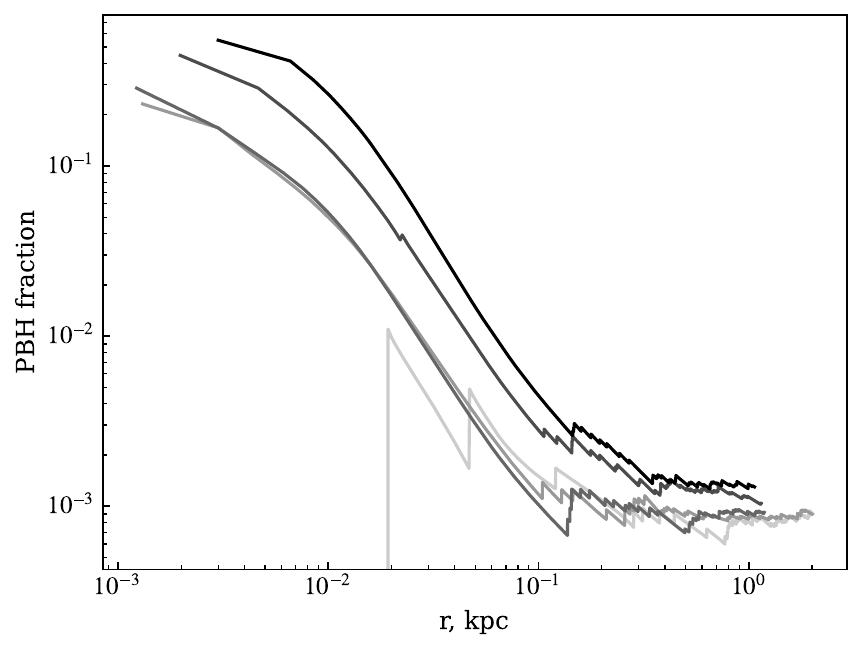}
    \caption{PBH fraction in a sphere of radius $ r $ as a function of time: the color gradations from light to dark correspond to moments $ a = 0.08 $ ($ z = 11.5 $), $ a = 0.1 $ ($ z = 9 $), $ a = 0.15 $ ($ z = 5.7 $), $ a = 0.27 $ ($ z = 2.7 $), $ a = 0.5 $ ($ z = 1 $).}
    \label{fig:time}
\end{figure}

Figure~\ref{fig:time} shows how the PBH fraction changed over time in a model with $ f_0 = 0.001 $, $ m_{PBH} = 500 \, M_\odot $ ($ m_{PBH}/m_{DM} = 100 $): after 20 million years from the start of halo formation, i.e., only a few dynamical times, 4 PBHs were found in the center, and over 5 billion years, this number gradually increased to 14, leading to an increase in the fraction.

We believe that the main reason for the rapid movement of several PBHs into the center of the galactic halo is the collisions between minihalos with each other and with the forming galactic halo. As shown in \cite{BarnesHut86}, when self-gravitating clusters of particles with comparable masses collide, they lose all relative motion energy and merge, i.e., such collisions are inelastic. Observations of PBH trajectories showed that the minihalos collide with each other during the turbulent early stage of galaxy halo formation.

It is also easy to estimate the probability of minihalo collisions during the early stage of halo formation. According to Zel'dovich's approximation, the initial conditions (\ref{eq:init}) evolve in time such that within 20 million years from $ a = 0.08 $ to $ a = 0.1 $, about 40\% of the matter in the box manages to pass through a plane perpendicular to the X-axis in the middle of the box (see Figure~\ref{fig:xy}). With $ f_0 = 0.001 $, on average, 167 PBHs with mass $ m_{PBH} = 500 \, M_\odot $ ($ m_{PBH}/m_{DM} = 100 $) are present in the box, and in the region that has passed through the caustic, $ N_c = 0.4 \times 167 = 67 $ PBHs are present. The caustic surface area per PBH is $ S_1 = L_{box}^2/N_c $. If the collision cross-section of a minihalo is $ \sigma_c = \pi r_h^2 $, where $ r_h $ is the minihalo radius, then, according to the statistics of a two-dimensional Poisson process, the probability of a minihalo collision is $ 1 - \exp(-\sigma_c/S_1) $. We measured the virial radii of the minihalos and found $ r_h = 0.35 $~kpc for the considered model. For this radius, we obtain a probability of 13\%. According to our observations, about 10\% of the minihalos collided with other minihalos during the galaxy halo formation.

Minihalo collisions lead to a rapid loss of energy and angular momentum, which significantly distinguishes DM fragmented into clusters (minihalos) from ordinary diffuse DM. In diffuse DM, energy loss occurs more slowly due to particle motion in the time- and space-varying gravitational potential created by other particles. Therefore, minihalos ``fall'' more quickly into the center of the forming halo than ordinary DM. As a result, in the region with a radius of 10~pc, 90\% of the DM particles originated from minihalos associated with PBHs.

The further gradual increase in the PBH fraction in Figure~\ref{fig:time} is apparently due to dynamical friction between PBHs and PDM. Minihalos around PBHs, due to their early formation, have a high density in the central region, and their density profile is close to a power law with a slope of $ r^{-9/4} $ \cite{1985ApJS...58...39B,Inman_19,Pilipenko22}. When they enter the galactic halo and during the turbulent initial stage of halo formation, minihalos lose some of their mass from the periphery, but the central, densest parts of the minihalos remain. The virial mass of minihalos outside the galactic halo at $ a = 0.5 $ is $ 150 \pm 20 \, m_{PBH} $, while in the center of the galactic halo, it is $ 10 - 30 \, m_{PBH} $. The characteristic radius of the minihalos containing half of their mass is on average $ 200(m_{PBH}/100M_\odot)^{1/3} $~pc outside the halo and $ 10 - 40 (m_{PBH}/100M_\odot)^{-1/3} $~pc inside the halo at distances greater than 40~pc from the center. Compact minihalos that survive inside the galactic halo will experience dynamical friction almost analogously to point particles with corresponding masses until they approach the center at a distance on the order of their radius. The characteristic time of dynamical friction is estimated using the formula from \cite{BT1987}:
\begin{equation}\label{eq:df}
    \tau = \frac{v^3}{4\pi \ln \Lambda G^2 \rho M_h},
\end{equation}
where $ M_h = 30 m_{PBH} $ is the minihalo mass, $ v $ is its characteristic velocity (also equal to the characteristic velocity of PDM), $ \ln \Lambda \approx 5 $ is the Coulomb logarithm. For $ m_{PBH} = 500 \, M_\odot $ and a characteristic velocity in the center of the halo of 5~km/s, we obtain $ \tau \approx 700 $ million years, which indicates the significant role of dynamical friction in this problem on timescales of over a billion years.

We performed similar measurements of the PBH fraction for models in which PBHs were introduced into the system later, at $ a_\mathrm{init} = 3 \times 10^{-4} $, which resulted in a 3-fold decrease in the minihalo mass. This led to a slight change in the PBH fraction. In the model with late PBH addition, at $ a_\mathrm{init} = 0.043 $ ($ z_\mathrm{init} = 22 $), just before halo formation, the PBH fraction in the center decreased by a factor of 6 compared to models with earlier PBH additions. This demonstrates the importance of minihalos for the condensation of PBHs in the center of the halo.

Finally, we also checked the change in the dark halo density profile, as previous studies \cite{Zhu18,Boldrini20} have shown that over long times, dynamical friction of PBHs can destroy the dark matter cusp. We observe a small, 2-fold, decrease in the DM density at the center compared to the NFW profile for some models. However, this does not lead to a significant change in the velocity dispersion of the particles, which varies within 0.5~km/s. Therefore, such a profile change is unlikely to be detectable in observations or responsible for solving the ``cusp problem'' (see, for example, the discussion of the problem in \cite{2021Galax..10....5B}). We leave the question of profiles for future investigation, in particular, further calculations with a longer duration (up to $ a = 1$ instead of $a=0.5$) are required.

\textbf{4. PBH fraction constraints.}
We approximate the dependence of the PBH fraction in a sphere of radius 10~pc, $f_{10}$, on the parameters: initial PBH fraction $f_0$, PBH mass $m_{PBH}$, and system scale $s$ with the following formula, which has an accuracy better than 30\% for all our simulations:
\begin{equation}\label{eq:f10}
    f_{10} = 43 f_0 \left( \frac{m_{PBH}}{100\;\mathrm{M}_\odot} \right)^{1/2}.
\end{equation}
The dependence on the system scale $s$ has diminished, but the system scale should still be taken into account when determining the applicability of formula (\ref{eq:f10}). Indeed, our numerical models cover the mass range of $100-5000$~M$_\odot$ at $s=1$. Considering the accuracy of formula (\ref{eq:f10}), we believe that it can be extrapolated by an order of magnitude in both directions, i.e., $10-5\times10^4$~M$_\odot$ while maintaining accuracy at a factor of 2. Now, let us remind that in our numerical calculations, we fix the mass ratio of PBHs to test particles of PDM, thereby fixing the ratio of PBH mass to the virial mass of the halo. Therefore, the mass range of our calculations scales as
\begin{equation}\label{eq:mlim}
    10\;\mathrm{M}_\odot < m_{PBH} \left( \frac{5\times10^7\;\mathrm{M}_\odot}{M_v} \right) < 5\times10^4\;\mathrm{M}_\odot.
\end{equation}
As for the initial PBH fraction range $f_0$, which in our calculations ranged from $10^{-4}$ to $10^{-3}$, we also consider it possible to extrapolate the results down to $f_0=10^{-5}$, provided that the number of PBHs in the region of interest is greater than 1. The average number of PBHs in the central cluster is approximated by the expression:
\begin{equation}\label{eq:npbh}
    N_{PBH} = 3\times10^4 f_0 \left( \frac{100\;\mathrm{M}_\odot}{m_{PBH}} \frac{M_v}{5\times10^7\;\mathrm{M}_\odot} \right)^{1/2} .
\end{equation}
Increasing the PBH fraction $f_0$ above $10^{-3}$ will probably cause a change in the density profile, as described in \cite{Boldrini20}, so formula (\ref{eq:f10}) will no longer be applicable.

\begin{figure}
    \centering
    \includegraphics[width=\linewidth]{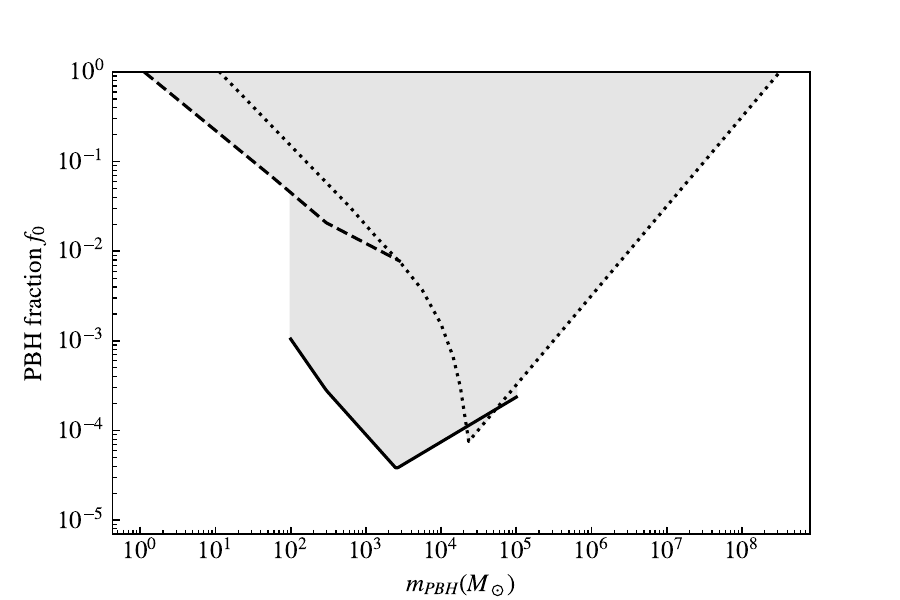} 
    \caption{Constraints on the cosmological PBH fraction, $f_0$, from the dynamics of dwarf galaxies. Solid line: results from this work, dashed line: results from \cite{Koulen24}, dotted line: results from \cite{Graham24a}.}
    \label{fig:main}
\end{figure}

Now we can apply formula (\ref{eq:f10}) to estimate the improvement in constraints on the PBH fraction $f_0$ obtained in the works \cite{Koulen24, Graham24a}. In \cite{Koulen24}, the star cluster at the center of the Eridanus II galaxy was considered. We assume that if the PBH fraction $f_{10}$ exceeds the value shown in Figure 5 of \cite{Koulen24}, the star cluster would have been destroyed by now. The virial mass of the galaxy in \cite{Koulen24} was taken to be $M_v=2.5\times10^8$~M$_\odot$, which allows us to set the PBH mass range from equation (\ref{eq:mlim}). We proceed in a similar way with the results from \cite{Graham24a}, where the star cluster in the Segue I galaxy with a virial mass of $M_v=10^9$~M$_\odot$ was considered. At the same time, we take results that do not account for PBH migration, as we account for it in our work. The combined result, also considering the condition $N_{PBH}>1$ from formula (\ref{eq:npbh}), is shown in Figure \ref{fig:main}.

\textbf{5. Conclusions.}
Our study of the formation of dark matter halos with a small fraction of PBHs showed that the inclusion of the ``dress'' or minihalos of DM around PBHs is extremely important: minihalos collide with each other, leading to the rapid formation of a PBH cluster at the center of dwarf galaxy halos. Thus, the fraction of PBHs in the DM content in the centers of dwarf galaxies increases by tens to hundreds of times compared to the cosmological fraction, depending on the PBH mass, see formula (\ref{eq:f10}). This fact allows us to update the known constraints on the PBH fraction from the literature, reducing them by about 2 orders of magnitude in the PBH mass range from 100 to 100 thousand M$_\odot$, as shown in Figure \ref{fig:main}.

However, this result cannot yet be considered final, as several processes occurring in real galaxies were not taken into account in our calculations. First, we did not consider small-scale cosmological structures — low-mass halos of PDM, which should have formed before the main halo. The spectrum of matter density perturbations in cosmology has not yet been studied on sufficiently small scales to unambiguously determine the number of these small structures (see, for example, \cite{Tkachev24}). We believe that collisions of minihalos containing PBHs with halos of comparable masses made of PDM will lead to an even greater enhancement of PBH migration to the center of the galactic halo.

Another important process for obtaining constraints from observations is the evolution of the baryonic matter component: in the works we relied on \cite{Koulen24, Graham24a}, the star cluster was initially assumed to be stationary. However, in our calculations, the PBH cluster forms quickly, within a few million years at the beginning of the halo formation. Therefore, the star formation process from gas must have been occurring already in the presence of the PBH cluster. This cluster may, among other things, contribute to star formation, see, for example, \cite{Dolgov17}. Solving these and other issues is possible in the future but will require significant computational resources.

Due to the scalability of our results, they may also be useful for tasks involving clusters made of low-mass PBHs, with PBH masses less than stellar masses, for example, see the work \cite{Belotsky24}. Our formula (\ref{eq:npbh}) allows estimating the number of PBHs in such clusters, given the mass function of the DM halos.

\textbf{Acknowledgements.}
The authors express their sincere gratitude to S.Yu. Dedikov and the team of administrators supporting the computational resources of the ASC LPI. The authors also thank P.B. Ivanov for valuable comments.

\textbf{Funding.}
This work was funded by the budget of the P.N. Lebedev Physical Institute of the Russian Academy of Sciences. No additional grants were received for the conduct or management of this specific research.

\textbf{Conflicts of Interest.} 
The authors declare that they have no conflicts of interest.


\bibliographystyle{ieeetr}
\bibliography{class.bib}

\end{document}